\documentclass[fleqn,review,12pt,sort&compress]{elsarticle}

\usepackage[small,justification=centerlast]{caption}
\usepackage{graphicx}
\usepackage{bm}
\usepackage{mathrsfs}
\usepackage[latin1]{inputenc}
\usepackage{stmaryrd}
\usepackage{array}
\usepackage{psfrag}
\usepackage{dsfont}
\usepackage{epsfig}
\usepackage{titletoc}
\usepackage{float}   
\usepackage{wrapfig} 
\usepackage{amsmath}
\usepackage[latin1]{inputenc}
\usepackage{amsmath}\usepackage{amssymb,stmaryrd}
\usepackage{mathrsfs}
\usepackage{array}
\usepackage{graphicx}
\usepackage{psfrag}
\usepackage{dsfont}
\usepackage{epsfig}
\usepackage{cancel}
\usepackage[dvips]{feynmp}
\usepackage{upgreek}
\usepackage{subfig}
\usepackage{tabularx}
\usepackage[dvipsnames]{xcolor}
\usepackage{units}
\usepackage{booktabs}
\usepackage[bottom]{footmisc}

\usepackage{textfit} 

\usepackage{hyperref}

\def\twiddle{\lower4pt\hbox{\hskip-0pt{$\widetilde{}$}}}
\def\m@th{\mathsurround=0pt}
\def\cmapstochar{\mathrel{\rlap{
  \lower0.1pt\hbox{\hskip-1.75pt{$\mapstochar$}}}
  \raise0pt\hbox{\hskip2.5pt{$\twiddle$}}}}
\def\notsimfill{$\m@th\cmapstochar$}
\def\scroodle#1{\vbox{\ialign{##\crcr\notsimfill\crcr
  \noalign{\kern-4pt\nointerlineskip}
   $\hfil\displaystyle{#1}\hfil$\crcr}}}

\renewcommand{\eqref}[1]{\mbox{Eq.~(\ref{#1})}}
\newcommand{\tabref}[1]{\mbox{Tab.~\ref{#1}}}
\newcommand{\figref}[1]{\mbox{Fig.~\ref{#1}}}
\newcommand{\secref}[1]{\mbox{Sec.~\ref{#1}}}

\begin{document}

\begin{frontmatter}

\title{Fermionic Lorentz violation and its implications for \\ interferometric gravitational-wave detection}

\author{M. Schreck\corref{cor1}}
\cortext[cor1]{Corresponding Author}
\ead{marco.schreck@ufma.br}

\address{Departamento de F\'{i}sica, Universidade Federal do Maranh\~{a}o \\
65080-805, S\~{a}o Lu\'{i}s, Maranh\~{a}o, Brazil}

\begin{abstract}
The recent direct detection of gravitational waves reported by Advanced LIGO has inspired the current article.
In this context, a particular Lorentz-violating framework for classical, massive particles is the focus. The latter
is characterized by a preferred direction in spacetime comprised of {\em CPT}-odd components with mass dimension 1.
Curvature effects in spacetime, which are caused by a propagating gravitational wave, are assumed to deform the otherwise
constant background field. In accordance with spontaneous Lorentz violation, a particular choice for the vector field
is taken, which was proposed elsewhere. The geodesic equations for a particle that is subject to this type
of Lorentz violation are obtained. Subsequently, their numerical solutions are computed and discussed. The particular
model considered leads to changes in the particle trajectory whose impact on interferometric gravitational-wave experiments
such as LIGO will be studied.
\end{abstract}

\end{frontmatter}

\newpage
\setcounter{equation}{0}
\setcounter{section}{0}
\renewcommand{\theequation}{\arabic{section}.\arabic{equation}}

\section{Introduction}

The recently reported direct detection of gravitational waves by Advanced LIGO \cite{Abbott:2016blz} counts as one of the most important discoveries in physics.
In his seminal 1916 paper \cite{Einstein:1916}, Einstein was able to show that a linearized version of his field equations exhibits wave-like solutions.
Gravitational waves are generated whenever masses in a gravitational system are accelerated such that certain derivatives of multipole moments of the system
are nonvanishing \cite{Moore:2013}. For example, this concerns nonspherical supernova explosions, rotations of binary pulsars, and black-hole mergers.
After all, the source of the signal GW150914 detected by Advanced LIGO is considered to have been the merger of two black holes.

Einstein thought that gravitational waves were a mere theoretical manifestation of local Lorentz invariance in General Relativity, but he did not have any
confidence in their experimental discovery. However, even before their direct detection by Advanced LIGO, the majority of the physics community had already
believed in gravitational waves. Observing the Hulse-Taylor binary over several decades confirmed that the measured energy loss of this system was in
excellent agreement with calculations based on gravitational waves in General Relativity \cite{Taylor:2005,Weisberg:2016jye}. Therefore, this result had
already been considered as indirect evidence for such waves.

Gravitational waves are tiny ripples that propagate through spacetime and generate strain. Typical
amplitudes far away from the source are smaller than one part in $10^{20}$. Hence, their direct detection is a significant technological challenge. The signal
GW150914 was only possible to be detected by an interferometer whose arm length is several kilometers and which, apart from that, involves various
sophisticated setups to reduce vibrations, thermal noise, etc. Advanced LIGO is a Michelson interferometer with Fabry-Perot arm cavities \cite{Fritschel:2015}.
This setup allows for sending light back and forth multiple times to increase its intensity.

The direct detection of gravitational waves can be considered as a lucky chance for the field of Lorentz symmetry violation. It will allow for obtaining
new constraints on Lorentz violation in the gravity sector as well as for massive elementary particles. Few days after the announcement by Advanced LIGO
the first paper appeared tightly bounding Lorentz violation in the pure-gravity sector \cite{Kostelecky:2016kfm} of the Standard-Model Extension
(SME). The SME is the most general Lorentz-violating framework. It comprises all possible Lorentz-violating
operators of the Standard Model of elementary particles \cite{Colladay:1998fq,Kostelecky:2009zp,Kostelecky:2011gq,Kostelecky:2013rta} and General
Relativity \cite{Kostelecky:2003fs}. Its minimal version \cite{Colladay:1998fq} involves all power-counting renormalizable terms, whereas the nonminimal
SME \cite{Kostelecky:2009zp,Kostelecky:2011gq,Kostelecky:2013rta} contains all operators of mass dimension $\geq 5$. For a review on various aspects of
Lorentz violation and the SME, consult \cite{Tasson:2014dfa}.

Since Advanced LIGO announced their result there has been a couple of papers on tests of alternative gravity theories using gravitational waves, amongst them
\cite{Collett:2016dey,Konoplya:2016pmh,Yagi:2016jml,Yunes:2016jcc}. The motivation for this paper is along the same lines. In particular, we intend to
consider how Lorentz violation in the matter sector could affect the detection of gravitational waves. To do so, a simple model for a Lorentz-violating
background field, which is deformed by a propagating wave, is introduced. The model
is based on the Lagrangian of a classical, relativistic, pointlike particle related to the SME $a$ coefficients, which are {\em CPT}-odd and of mass
dimension one. For the past few years research has been performed on how a classical-particle description can be obtained from the field theory language
of the SME. This branch of research was initiated by the paper \cite{Kostelecky:2010hs} covering many cases of the minimal fermion sector. Subsequent works complemented
our knowledge of the minimal fermion sector \cite{Kostelecky:2011qz,Kostelecky:2012ac,Colladay:2012rv,Schreck:2014ama,Russell:2015gwa,Schreck:2016jqn}
where some articles discussed the nonminimal fermion sector \cite{Schreck:2014hga,Schreck:2015seb} as well as the photon sector \cite{Schreck:2015dsa}.
These classical Lagrangians are particularly suitable when describing Lorentz-violating particles on curved manifolds, which is why we will use this method
here.

The paper is organized as follows. In \secref{sec:gravitational-waves-lorentz-violation} the most important properties of gravitational waves are
reviewed. In this context, the particular classical Lagrangian, which forms the foundation of the studies, is introduced and coupled to the metric perturbation
induced by the gravitational wave. In \secref{sec:models} we consider a particular model based on this Lagrangian. It is characterized by a
vector field that violates Lorentz invariance spontaneously. Based on results obtained in
the context of the gravitational SME, the geodesic equations for a massive particle are derived. Subsequently, this set of differential equations is
solved numerically in \secref{eq:solution-equations-of-motion}. First of all, a solution is computed for hypothetical numerical parameters to allow
for a convenient and transparent discussion of the behavior of a massive particle under the combined influence of the gravitational
wave and the Lorentz-violating background field. Section \ref{sec:observable-signals} is dedicated to exploring what impact this behavior has on
realistic interferometric setups such as Advanced LIGO. Finally, the results are concluded on in \secref{sec:conclusion}. In
\ref{sec:remarks-finsler-geometry} we discuss explicit Lorentz violation in a gravitational-wave metric from the perspective of Finsler geometry.
Thereby, an interesting observation is made with regards to the geodesic equations. Throughout the paper, natural units are used with $c=\hbar=1$ unless
otherwise stated.

\section{Gravitational waves and Lorentz-violating classical particles}
\label{sec:gravitational-waves-lorentz-violation}

Gravitational waves are solutions of the linearized Einstein equations. To solve those equations it is reasonable to separate the curved spacetime metric
$g_{\mu\nu}$ into the Minkowski metric $\eta_{\mu\nu}$ and a piece $h_{\mu\nu}$ that comprises all curvature effects:
\begin{equation}
\label{eq:curved-spacetime-metric}
g_{\mu\nu}=\eta_{\mu\nu}+h_{\mu\nu}\,.
\end{equation}
The Einstein equations can be linearized in the weak-field limit, which translates to the claim that $h_{\mu\nu}\ll \eta_{\mu\nu}$. It is possible
to simplify the resulting linearized equations drastically on the foundation of diffeomorphism invariance of General Relativity. This means that there
is a set of gauge transformations that do not change the Einstein equations. A particular gauge choice, which is a combination of the Lorenz gauge and
the transverse-traceless gauge \cite{Moore:2013}, leads to a wave equation for $h_{\mu\nu}$ that has the following solution:
\begin{subequations}
\label{eq:metric-perturbation-gravitational-wave}
\begin{align}
h_{\mu\nu}&=\left(A_+\Pi^{(+)}+A_{\times}\Pi^{(\times)}\right)_{\mu\nu}\cos(k\cdot X)\,, \displaybreak[0] \\[2ex]
\label{eq:polarization-matrices}
\Pi^{(+)}_{\mu\nu}&=\begin{pmatrix}
0 & 0 & 0 & 0 \\
0 & 1 & 0 & 0 \\
0 & 0 & -1 & 0 \\
0 & 0 & 0 & 0 \\
\end{pmatrix}_{\mu\nu}\,,\quad \Pi^{(\times)}_{\mu\nu}=\begin{pmatrix}
0 & 0 & 0 & 0 \\
0 & 0 & 1 & 0 \\
0 & 1 & 0 & 0 \\
0 & 0 & 0 & 0 \\
\end{pmatrix}_{\mu\nu}\,.
\end{align}
\end{subequations}
The latter has been restricted to its real part. Here $k_{\mu}$ is the wave vector, $X^{\mu}=(t,\mathbf{x})^{\mu}$ are the coordinates, and $A_+$, $A_{\times}$ are called
the strain amplitudes. Recall that the amplitude of an electromagnetic wave is characterized by its electric and magnetic field strength vectors whose
magnitudes change periodically. An analog vectorial picture is not valid for gravitational waves since the latter are perturbations of the metric, which
is a two-tensor. Therefore, gravitational waves can be understood as metric perturbations whose amplitudes $A_+$, $A_{\times}$ vary periodically. When a
gravitational wave propagates through spacetime this leads to what is known as strain. In analogy to electromagnetic waves, these tiny ripples in spacetime
can have two different polarizations, which are denoted as (+) and ($\times$). Each is described by one of the particular matrices $\Pi^{(+)}$ and $\Pi^{(\times)}$,
respectively, which are to be found in \eqref{eq:polarization-matrices}.
Both matrices are traceless due to the choice of the transverse-traceless gauge. The notation for the polarizations results from the effect that they
have on a ring of massive particles. For the (+) polarization this ring is distorted to an ellipse whose eccentricity changes periodically. Both
semiaxes are oriented to point along the horizontal and vertical axes, respectively, of a Cartesian coordinate system. For the $(\times)$ polarization
the behavior of the ring is similar, but the semiaxes of the ellipse are rotated by $45^{\circ}$.

In this paper the gravity sector itself is assumed to be standard, i.e., the form of a gravitational wave is given by
\eqref{eq:metric-perturbation-gravitational-wave} with the usual dispersion relation $\omega=|\mathbf{k}|$. Therefore, gravitational waves shall
propagate with the speed of light. In \cite{Kostelecky:2016kfm} modifications of the latter dispersion relation in the pure-gravity sector
of the SME are considered leading to tight constraints on the nonminimal coefficients based on GW150914. In the current article, Lorentz
violation is taken to reside in the fermion sector. The general Lagrangian containing all minimal and nonminimal operators is stated in Eq.~(1) of
\cite{Kostelecky:2013rta} including the subsequent definitions.
We consider an observer frame with nonzero minimal $a$ coefficients only, i.e., the fermion action reads as follows \cite{Kostelecky:2013rta}:
\begin{equation}
\label{eq:sme-fermion-a-coefficients}
S=\int \mathrm{d}^4X\,\mathcal{L}\,,\quad \mathcal{L}=\frac{1}{2}\overline{\psi}(\gamma^{\mu}\mathrm{i}\partial_{\mu}-m_{\psi}\mathds{1}_4-a_{\mu}\gamma^{\mu})\psi+\text{H.c.}
\end{equation}
Here $\psi$ is a spinor field, $\gamma^{\mu}$ are the standard Dirac matrices based on the Clifford algebra, $m_{\psi}$ is the fermion mass, and
$\mathds{1}_4$ is the unit matrix in spinor space. Lorentz violation is encoded in the nonzero observer four-vector $a_{\mu}$. Note that the
Lagrange density, as it stands, is defined in Minkowski spacetime. Thereby, we use the signature $(+,-,-,-)$ of the Minkowski metric $\eta_{\mu\nu}$
in contrast to the majority of the literature
related to General Relativity. To study the effect of a gravitational wave on
particles subject to such a particular type of Lorentz violation it is reasonable to resort to an analog, classical description. Indeed, there is
a procedure that allows for mapping a field theory such as \eqref{eq:sme-fermion-a-coefficients} to a classical-particle description. For the SME
the relevant tools were set up in \cite{Kostelecky:2010hs} with further studies performed in a subsequent series of papers
\cite{Colladay:2012rv,Schreck:2014ama,Russell:2015gwa,Schreck:2016jqn,Schreck:2014hga,Schreck:2015seb,Schreck:2015dsa}. The Lagrangian describing a
classical, relativistic, pointlike particle subject to the $a$ coefficients was derived in \cite{Kostelecky:2010hs}. It is given as follows:
\begin{equation}
\label{eq:lagrangian-a}
L^{(a)}=-m_{\psi}\sqrt{u^2}-a\cdot u\,,
\end{equation}
where $u^{\mu}=(u^0,\mathbf{u})^{\mu}$ is the four-velocity of the classical particle. The Lagrangian $L^{(a)}$ is defined in Minkowski spacetime, i.e., each pair of
four-vectors is contracted with $\eta_{\mu\nu}$ and the background vector $a_{\mu}$ is taken to be constant. The model considered will be based
on this particular Lagrangian for several reasons. First, it is the simplest one obtained in the context of the SME, which makes its study practical
from a technical point of view. Second, the $a$ coefficients are particularly interesting in gravity since possible effects from these coefficients
cannot be measured in Minkowski spacetime using only a single fermion species. The reason is that the $a$ coefficients can be eliminated by a field
redefinition, which is why they can, in principle, be very large in Minkowski spacetime~\cite{Kostelecky:2008in}. This changes in a gravitational
background where a global field redefinition is not possible. With the procedure of minimal coupling, the Lagrangian of \eqref{eq:lagrangian-a}
is promoted to a curved spacetime. Minimal coupling means to replace all occurrences of the Minkowski metric by the curved spacetime
metric $g_{\mu\nu}=g_{\mu\nu}(X)$ and promoting the constant $a_{\mu}$ to a function dependent on spacetime position,
cf.~\cite{Kostelecky:2011qz,Schreck:2015dsa,Schreck:2015seb}:
\begin{subequations}
\begin{align}
L^{(a)}[\eta_{\mu\nu},u^{\mu},a_{\mu}]&\mapsto \widetilde{L}^{(a)}[g_{\mu\nu}(X),u^{\mu},a_{\mu}(X)]\,, \\[2ex]
L^{(a)}[\eta_{\mu\nu},u^{\mu},a_{\mu}]&=-m_{\psi}\sqrt{\eta_{\mu\nu}u^{\mu}u^{\nu}}-a_{\mu}u^{\mu}\,, \\[2ex]
\label{eq:lagrangian-minimally-coupled}
\widetilde{L}^{(a)}[g_{\mu\nu}(X),u^{\mu},a_{\mu}(X)]&=-m_{\psi}\sqrt{g_{\mu\nu}(X)u^{\mu}u^{\nu}}-a_{\mu}(X)u^{\mu}\,.
\end{align}
\end{subequations}
The curved metric $g_{\mu\nu}$ is chosen according to \eqref{eq:curved-spacetime-metric} with the gravitational-wave perturbation of
\eqref{eq:metric-perturbation-gravitational-wave}. Integrating over the curve parameter of the particle yields the action, which corresponds
to Eq.~(68) of \cite{Kostelecky:2010ze} (when taking into account the different metric signature). The particle can also be considered as an
extended test mass, which consists of electrons, protons, and neutrons that are held together by a binding-energy contribution. Each individual
particle of such a test mass then moves with the same velocity \cite{Kostelecky:2010ze}. This description will be elaborated on in
\secref{sec:observable-signals}.

\section{Model}
\label{sec:models}

We will consider a gravitational wave propagating through free space far away from all massive bodies. The part of spacetime that is not affected
by the gravitational wave is assumed to be Minkowskian where the Lorentz-violating background field shall comprise constant controlling coefficients.
In a curved spacetime manifold, the notion of a constant vector or tensor field loses its meaning. Explicit Lorentz violation is then incompatible
with a curved spacetime manifold based on Riemannian geometry \cite{Kostelecky:2003fs}. The reason is that the modified energy-momentum conservation
law clashes with the geometrical Bianchi identities. Since explicit Lorentz violation is nondynamical, momentum cannot be transferred between
the particle and the background field leading to problems. The well-established possibility of circumventing this
no-go theorem is to resort to spontaneous Lorentz violation \cite{Bluhm:2008yt,Bluhm:2014oua}. In this context, a Lorentz-violating background field is taken to be dynamical
satisfying field equations, as well. Taking into account this geometric compatibility, a proper choice for $a_{\mu}$ is as follows
\cite{Kostelecky:2008in}:
\begin{subequations}
\label{eq:lorentz-violating-vectorfield}
\begin{align}
a_{\mu}&=\overline{a}_{\mu}+\scroodle{a}_{\mu}\,, \\[2ex]
\label{eq:lorentz-violating-vectorfield-scroodle-a}
\scroodle{a}_{\mu}&=\frac{\alpha}{2}\left(h_{\mu\nu}\overline{a}^{\nu}-\frac{1}{2}\overline{a}_{\mu}h^{\nu}_{\phantom{\nu}\nu}\right)\,.
\end{align}
\end{subequations}
The first contribution $\overline{a}_{\mu}$ is interpreted as the constant background in Minkowski spacetime and $\scroodle{a}_{\mu}$ is a small
fluctuation\footnote{For the latter, the notation has been adapted to match that of
\cite{Kostelecky:2010ze} for the fluctuation with lower indices.} that is induced by the gravitational wave passing by. In this case, $h_{\mu\nu}$ is the metric perturbation of the
gravitational wave and $\alpha$ is the coupling constant of $a_{\mu}$ to gravity. Note that $h^{\nu}_{\phantom{\nu}\nu}=0$ for the metric perturbation
in transverse traceless gauge, cf.~\eqref{eq:metric-perturbation-gravitational-wave}. Therefore, only the first term in $\scroodle{a}_{\mu}$ contributes
in this context. When the gravitational wave propagates through empty space the background field shall adapt to the metric perturbation
caused by the wave. The physical situation is illustrated in \figref{fig:propagating-gravitational-wave}.
\begin{figure}[t!]
\centering
\subfloat[]{\label{fig:propagating-gravitational-wave}\includegraphics[scale=0.3]{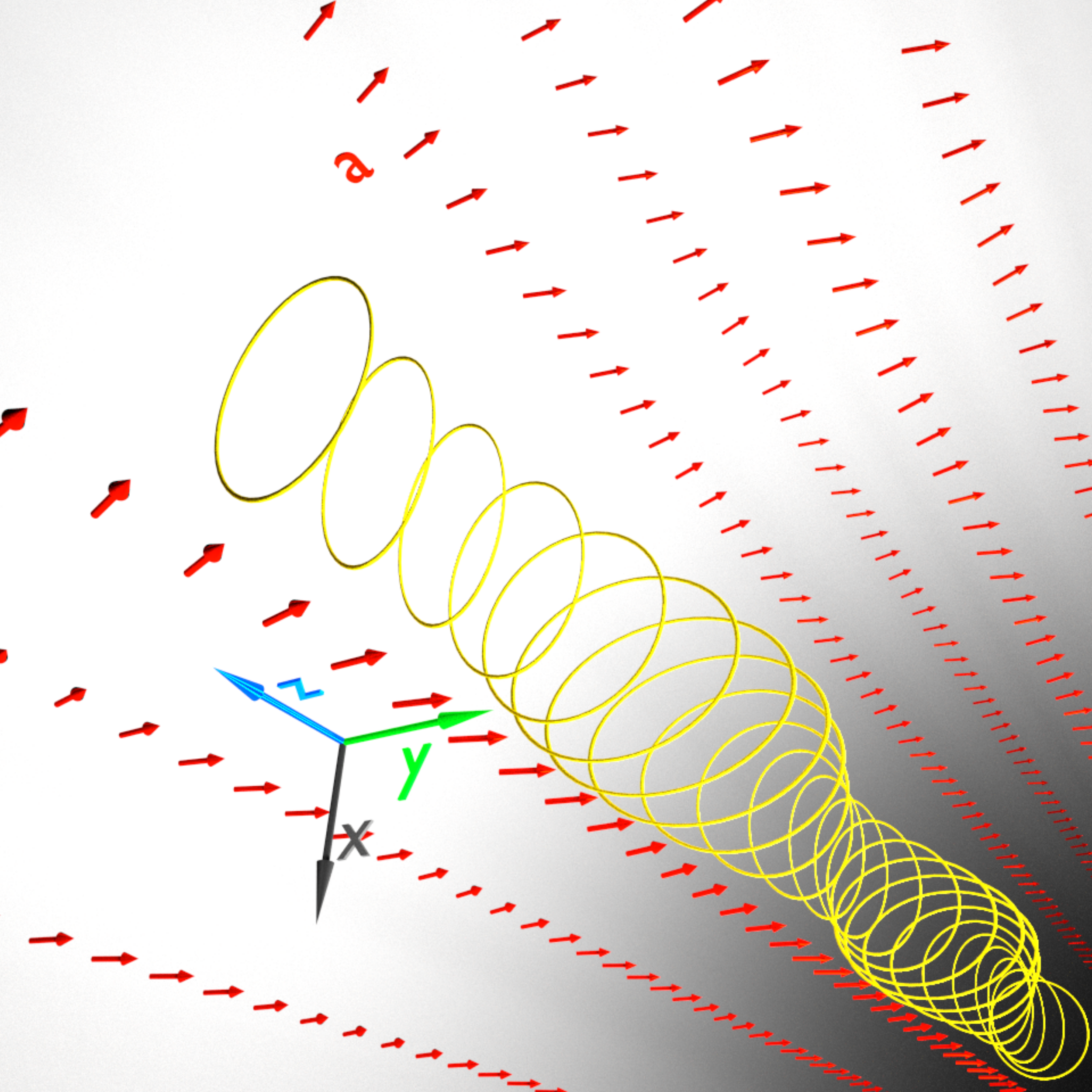}}\hspace{2cm}
\subfloat[]{\label{fig:gravitational-wave-vector-field}\includegraphics[scale=0.3]{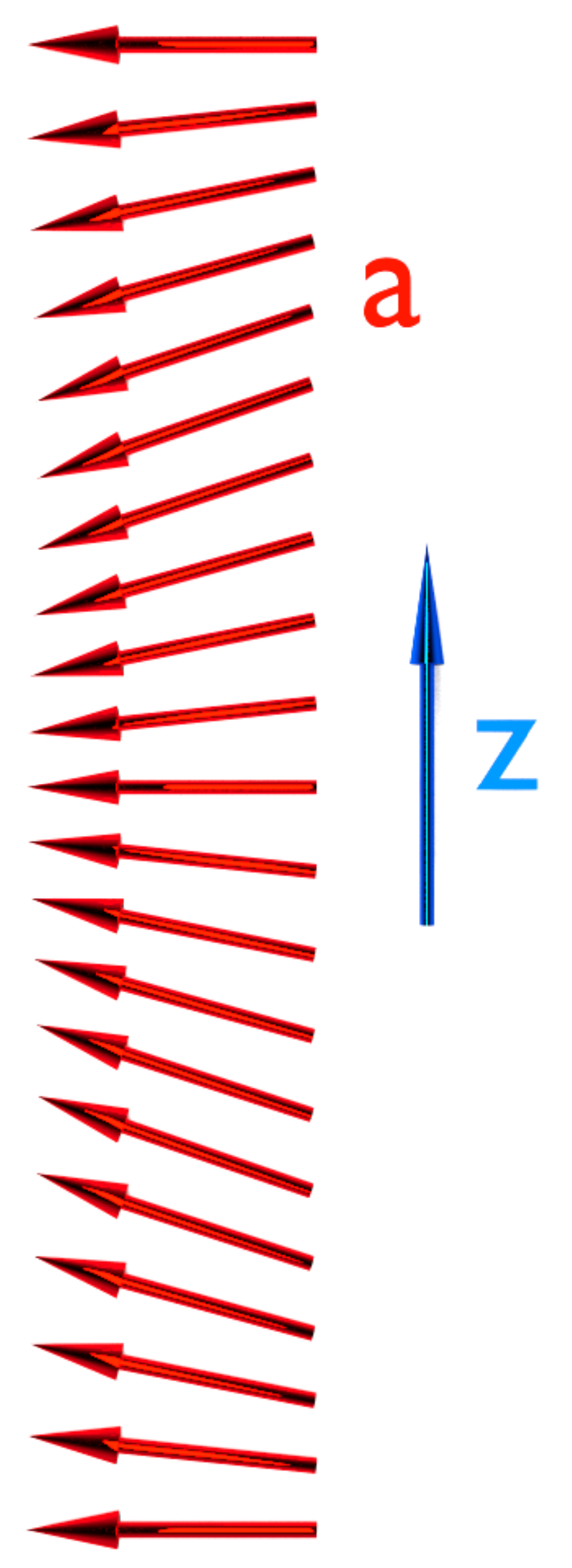}}
\caption{Panel~\protect\subref{fig:propagating-gravitational-wave} shows a propagating gravitational wave (yellow rings) that deforms the Lorentz-violating
background field $a_{\mu}$ (red arrows). The propagation direction points along the $z$ axis. For illustration purposes, only certain parts of the
Lorentz-violating vector field are visible. To better illustrate its modulation with the local gravitational-wave strain tensor a snapshot of a part
of the vector field is presented in \protect\subref{fig:gravitational-wave-vector-field}.}
\label{fig:propagating-gravitational-wave}
\end{figure}

The general form of the vector field given enables us to study the behavior of a
classical particle subject to Lorentz violation and a gravitational wave passing by. By observer Lorentz invariance, we choose a reference frame such
that the wave three-vector $\mathbf{k}$ points along the vertical $z$ axis. The wave four-vector is then given by $(k_{\mu})=(\omega,0,0,-k)$ with the
wave frequency $\omega$ and $k=|\mathbf{k}|=2\pi/\lambda$ where $\lambda$ is its wavelength. A classical particle is considered at the initial position
$x(0)=x_0$, $y(0)=y_0$ and $z(0)=0$ with its initial velocity components vanishing, $\dot{\mathbf{x}}(0)=\mathbf{0}$.

For a calculational purpose, we will make a couple of reasonable approximations. First, the source of the gravitational wave shall be far away from
the particle, which renders its amplitude strain minuscule: $A_{+,\times}\ll 1$. Second, all current constraints suggest that Lorentz violation is perturbative
leading to $|\scroodle{a}_{\mu}|/m_{\psi}\ll 1$. Recall that single-flavor experiments are not sensitive to $\overline{a}_{\mu}$.
Last but not least, the velocity changes of the particle are assumed to be much smaller than the speed of light, which is why $u^i\ll 1$.
The geodesic equations are obtained at first order in all small parameters.

\subsection{Geodesic equations}
\label{sec:geodesic-equations}

The Lorentz-violating background field must be chosen according to \eqref{eq:lorentz-violating-vectorfield} to be in accordance with
spontaneous Lorentz violation, whereupon its shape is already heavily restricted. The constant part is given by the general four-vector $\overline{a}_{\mu}$
with nonvanishing components. Furthermore, the gravitational wave is assumed to be in a superposition of both polarizations. The background field then reads
\begin{subequations}
\label{eq:choice-background-field-model}
\begin{align}
a_{\mu}(\Xi)&=\overline{a}_{\mu}+\begin{pmatrix}
0 \\
\breve{a}_1 \\
\breve{a}_2 \\
0 \\
\end{pmatrix}_{\mu}\Xi\,, \displaybreak[0]\\[2ex]
\label{eq:perturbation-coefficients}
\breve{a}_1&=-\frac{\alpha}{2}(A_+\overline{a}_1+A_{\times}\overline{a}_2)\,,\quad \breve{a}_2=\frac{\alpha}{2}(A_+\overline{a}_2-A_{\times}\overline{a}_1)\,, \displaybreak[0]\\[2ex]
\Xi&\equiv \cos(k\cdot X)\,.
\end{align}
\end{subequations}
The zeroth and third component of the spacetime-dependent part vanish. The first and second components involve both strain amplitudes but
the constant coefficients $\overline{a}_1$ and $\overline{a}_2$ only.
For this background field the geodesic equations are obtained from Eq.~(78) in \cite{Kostelecky:2010ze} by taking into account the first and the
last two terms on the right-hand side of the latter equation. They read as follows:
\begin{subequations}
\label{eq:geodesic-equations-model}
\begin{align}
\label{eq:geodesic-equation-t-component}
0&=\ddot{t}+\left(\dot{x}\frac{a_1'}{m_{\psi}}+\dot{y}\frac{a_2'}{m_{\psi}}\right)\omega\sin(k\cdot X)\,, \displaybreak[0]\\[2ex]
\label{eq:geodesic-equation-x-component}
0&=\ddot{x}+\left[A_+\dot{x}+A_{\times}\dot{y}+(1-\dot{z})\frac{a_1'}{m_{\psi}}\right]\omega\sin(k\cdot X)\,, \displaybreak[0]\\[2ex]
\label{eq:geodesic-equation-y-component}
0&=\ddot{y}+\left[A_{\times}\dot{x}-A_+\dot{y}+(1-\dot{z})\frac{a_2'}{m_{\psi}}\right]\omega\sin(k\cdot X)\,, \displaybreak[0]\\[2ex]
\label{eq:geodesic-equation-z-component}
0&=\ddot{z}+\left(\dot{x}\frac{a_1'}{m_{\psi}}+\dot{y}\frac{a_2'}{m_{\psi}}\right)\omega\sin(k\cdot X)\,,
\end{align}
\end{subequations}
where $a_{\mu}'\equiv \mathrm{d}a_{\mu}/\mathrm{d}\Xi$.
Since the particle is initially taken to be at rest and all Lorentz-violating coefficients and amplitude strains are assumed to be small, the
particle will move nonrelativistically. Hence, to a good approximation $\tau\approx t$ with the laboratory time $t$ and $u^0=\dot{t}\approx 1$,
$\mathbf{u}\approx \dot{\mathbf{x}}$. Possible Lorentz-violating effects at the source of the gravitational
wave will be neglected.

Note that in \ref{sec:remarks-finsler-geometry} the same equations of motion are derived in the context of explicit Lorentz violation within
Finsler geometry \cite{Antonelli:1993,Bao:2000}. The latter geometry goes beyond the quadratic restriction of line intervals on manifolds that
forms the base of Riemannian geometry. This change of the very foundation makes Finsler geometry capable of incorporating preferred directions
on curved manifolds. In \cite{Kostelecky:2003fs} it was proposed that explicit Lorentz violation could probably be described consistently in a
gravitational background by getting rid of the restrictions imposed by Riemannian geometry. The findings in \ref{sec:remarks-finsler-geometry}
may again indicate that spontaneous and explicit Lorentz violation in the setting of Finsler geometry are merely different descriptions of the same
physics, cf.~also \cite{Schreck:2015dsa}. However, whether or not explicit Lorentz violation is consistent within this generalized geometrical
framework is still an open problem to be solved.

\section{Solution of the equations of motion}
\label{eq:solution-equations-of-motion}

The geodesic equations (\ref{eq:geodesic-equations-model}) form a system of ordinary, coupled, nonlinear differential equations, which makes
their exact solution challenging. In all equations, the function $a_{\mu}[\cos(k\cdot X)]$ for a particular controlling coefficient has been
considered as arbitrary. In what follows, the choice given in \eqref{eq:choice-background-field-model}, will be made. First of all, the
equations of motion will be solved numerically based on a hypothetical set of numerical values for the parameters. By doing so, we will be in a
better position to discuss their solutions from a physical point of view.

\begin{figure}[t]
\centering
\subfloat[\label{fig:x-motion-model}]{\includegraphics[scale=0.45]{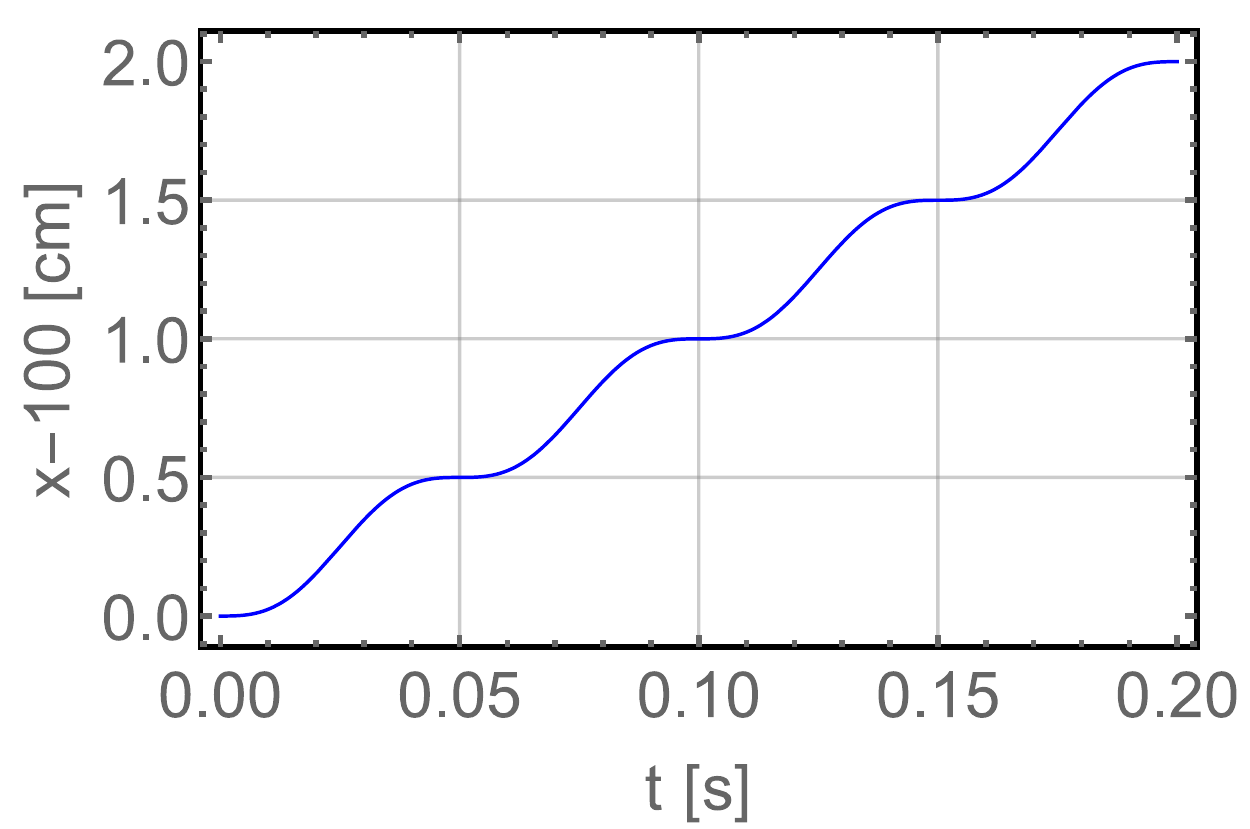}}
\subfloat[\label{fig:y-motion-model}]{\includegraphics[scale=0.45]{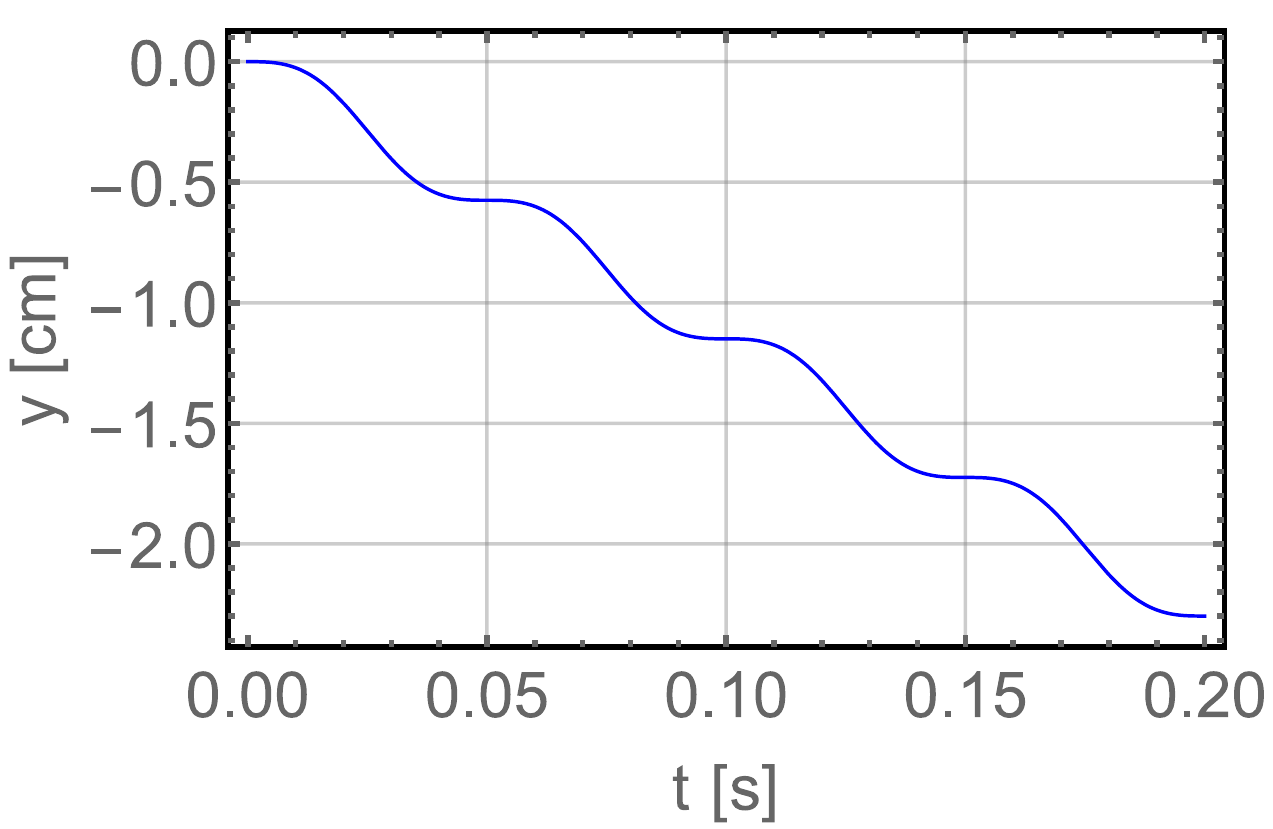}}
\subfloat[\label{fig:z-motion-model}]{\includegraphics[scale=0.45]{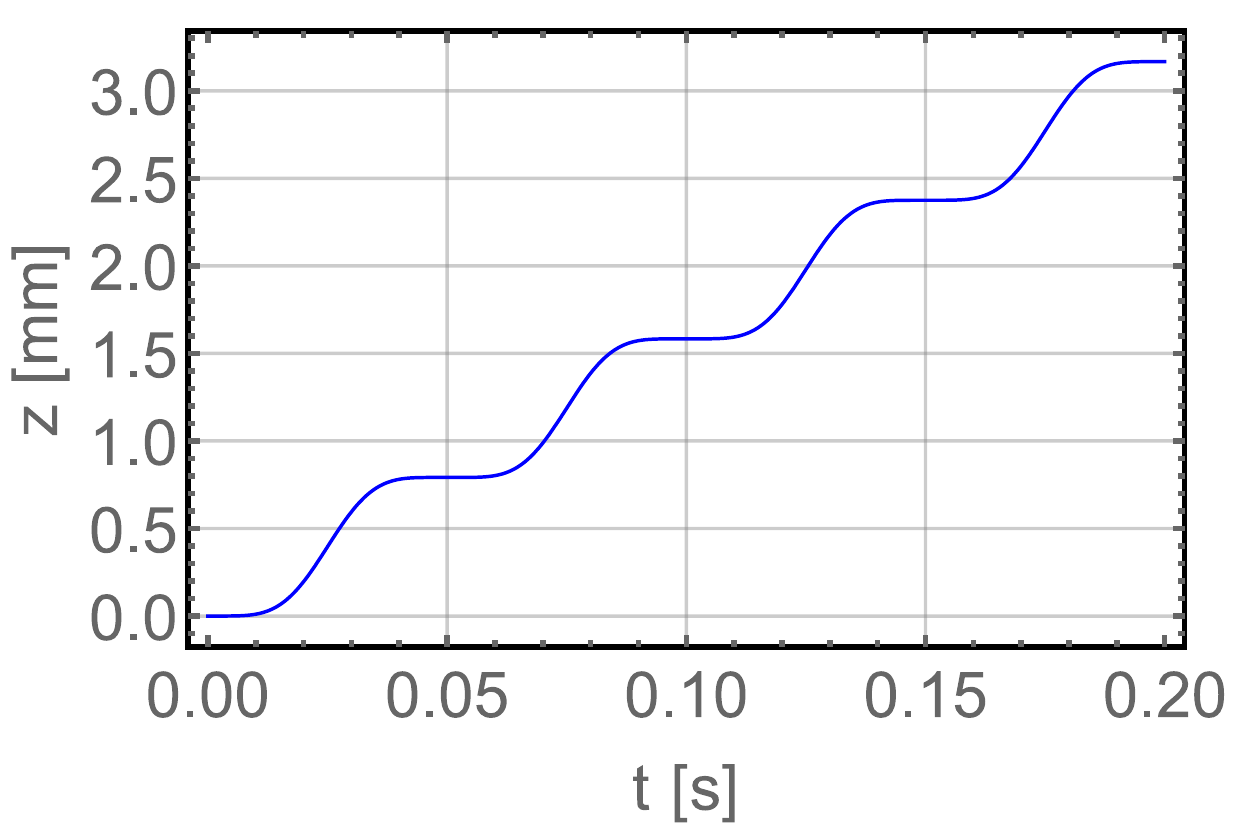}}
\caption{Law of motion for a pointlike particle along the $x$ \protect\subref{fig:x-motion-model}, the $y$ \protect\subref{fig:y-motion-model},
and the $z$ axis \protect\subref{fig:z-motion-model} based on Eqs.~(\ref{eq:geodesic-equations-model}) and the numerical values in the first
row of \tabref{tab:numerical-values}. The initial particle coordinates are $x_0=\unit[1]{m}$ and $y_0=z_0=0$. Note the
different units in \protect\subref{fig:z-motion-model} in comparison to the first two panels.}
\label{fig:motion-model}
\end{figure}

Furthermore, suitable initial-value conditions are necessary. These have partially been mentioned in \secref{sec:models}. Both proper time
and coordinate time are chosen such that $t=\tau=0$ when the gravitational wave hits the particle. Since the particle is taken to be at rest at
the beginning, it is supposed to start moving with a nonrelativistic speed, which is why $t\approx\tau$ to a good level of approximation. Hence, we
will work with the additional initial conditions $t(0)=0$ and $\dot{t}(0)=1$. The geodesic equations will be solved with the computer algebra
system \texttt{Mathematica} based on the values given in the first line of \tabref{tab:numerical-values}. The numerical solutions are shown in
\figref{fig:motion-model}.
\begin{table}[b]
\centering
\begin{tabular}{ccccccc}
\toprule
$\omega/(2\pi)$ [Hz] & $A_+$ & $A_{\times}$ & $\breve{a}_1/m_{\psi}$ & $\breve{a}_2/m_{\psi}$ & $T_0$ [s] \\
\midrule
20 & 0.1 & 0.1 & $-0.1$ & 0.1 & 0.2 \\
100 & $10^{-21}$ & 0.0 & 0.0 & 1.0 & 0.2 \\
\bottomrule
\end{tabular}
\caption{Numerical values for the gravitational-wave parameters, the dimensionless Lorentz-violating coefficients
(see \eqref{eq:choice-background-field-model}), and the integration time $T_0$ of the geodesic equations. The latter corresponds to the
duration of the event GW150914 detected by LIGO.}
\label{tab:numerical-values}
\end{table}

The behavior of the classical particle can be described as follows. The gravitational wave makes the particle move along all three spatial
directions. The particle starts traveling with the wave along the $z$ axis. The magnitude of its position increases monotonically and periodically
where its velocity oscillates. In time intervals, whose lengths stay constant, the particle gets accelerated first and decelerated after, i.e.,
the motion is periodic. As
long as the background field and the amplitude strains are small enough, the particle moves in phase with the wave. A qualitatively similar
motion occurs along the $x$ and $y$ directions, which just differs by the path length traveled. Since the equations of motion
are not symmetric under interchanging $x$ and $y$ the particle propagates slightly differently along the $x$ and $y$ axes. Also, particle
motion is more pronounced with the gravitational-wave strain and suppressed along the longitudinal direction.

Inserting the numerical solutions into the differential equations delivers an estimate of the numerical error $s$ of the solutions.
With this in mind, we integrate the force $F_i$ with respect to time to produce the momentum transfer $\Delta p_i$ to the particle along
the $i$-th axis. Carrying out the numerical integration over a single period $T=\unit[1/20]{s}$, gives
\begin{equation}
\Delta \mathbf{p}=m_{\psi} \int_0^T \mathbf{x}''(t)\,\mathrm{d}t\approx m_{\psi}\times \unit[\mathcal{O}(s)]{\frac{m}{s}}\,.
\end{equation}
This result is consistent with the numerical uncertainties originating from solving the equations of motion, which shows that
momentum is conserved over a full period. First, momentum is transferred from the gravitational wave to the particle
whereupon the latter is accelerated. Subsequently, the particle is decelerated again giving its momentum back to the gravitational wave.
Hence, the total momentum transfer over a single period is expected to vanish. The particle has moved a particular distance but stands
still again before the process repeats.

For the particle to move, both the gravitational wave and the Lorentz-violating background field are crucial. Without the vector field,
local Lorentz invariance is valid, and gravitational-wave physics is standard, which is why the position of a pointlike particle is not
affected. Without the gravitational wave, the vector field is homogeneous, and the geodesic in such a background is still a straight line.
Hence, a particle that rests in the lab frame will stay at rest despite the background field. The gravitational wave deforms the
background field and makes the latter inhomogeneous, which forces the particle to move. Note also that according to
\eqref{eq:lorentz-violating-vectorfield-scroodle-a}, there is no deformation $\scroodle{a}_{\mu}$ without nonzero strain amplitudes
$A_{+}$, $A_{\times}$ and nonzero constant background coefficients $\overline{a}_{\mu}$.

First of all, the particle propagates along the
$z$ axis as the gravitational wave induces a gradient of the vector field that points along that axis. Whether the particle moves
along the positive or negative $z$ axis is determined by the direction of the wave vector of the gravitational wave and the signs of the
controlling coefficients. Only a subset of the new terms in the equations of motion drives the propagation along the $z$ axis. What is needed
for this motion to occur is one of the terms including $a_i'$ in \eqref{eq:geodesic-equation-x-component} or (\ref{eq:geodesic-equation-y-component})
and one of the modified terms in \eqref{eq:geodesic-equation-z-component}.

\begin{figure}
\centering
\subfloat[\label{fig:ring-of-particles}]{\includegraphics[scale=1]{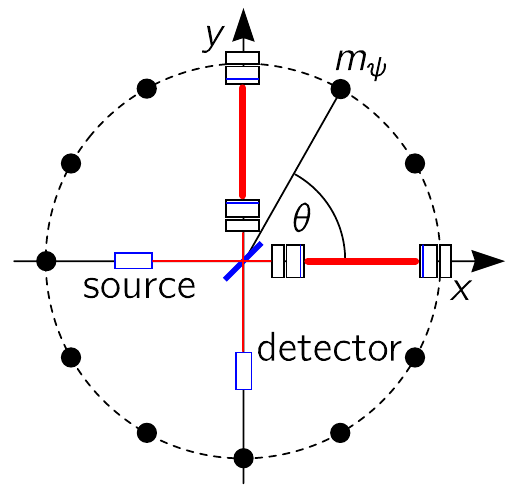}}\hspace{0.5cm}
\subfloat[\label{fig:ring-plus-polarization}]{\includegraphics[scale=0.4]{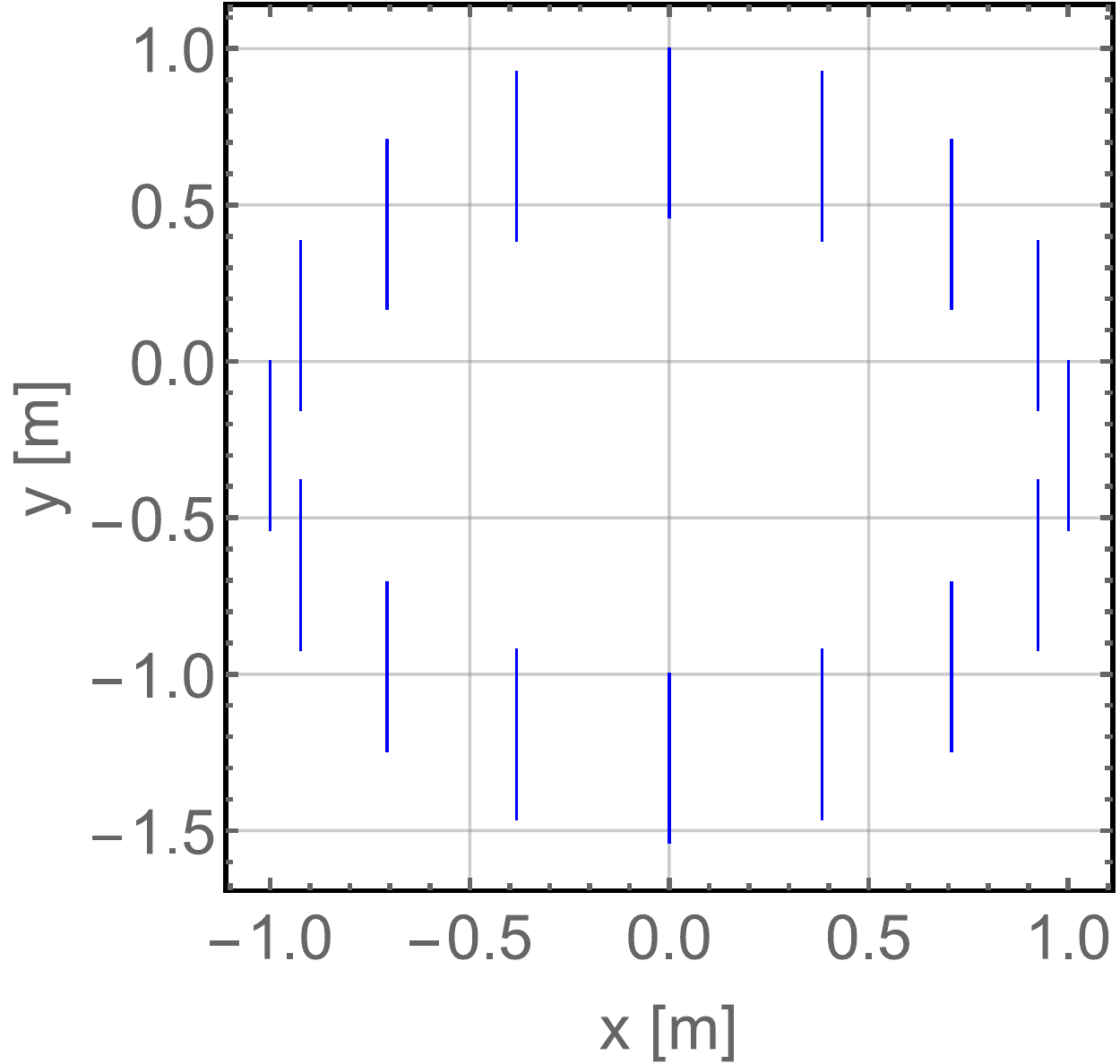}}\hspace{0.5cm}
\subfloat[\label{fig:ring-both-polarizations}]{\includegraphics[scale=0.4]{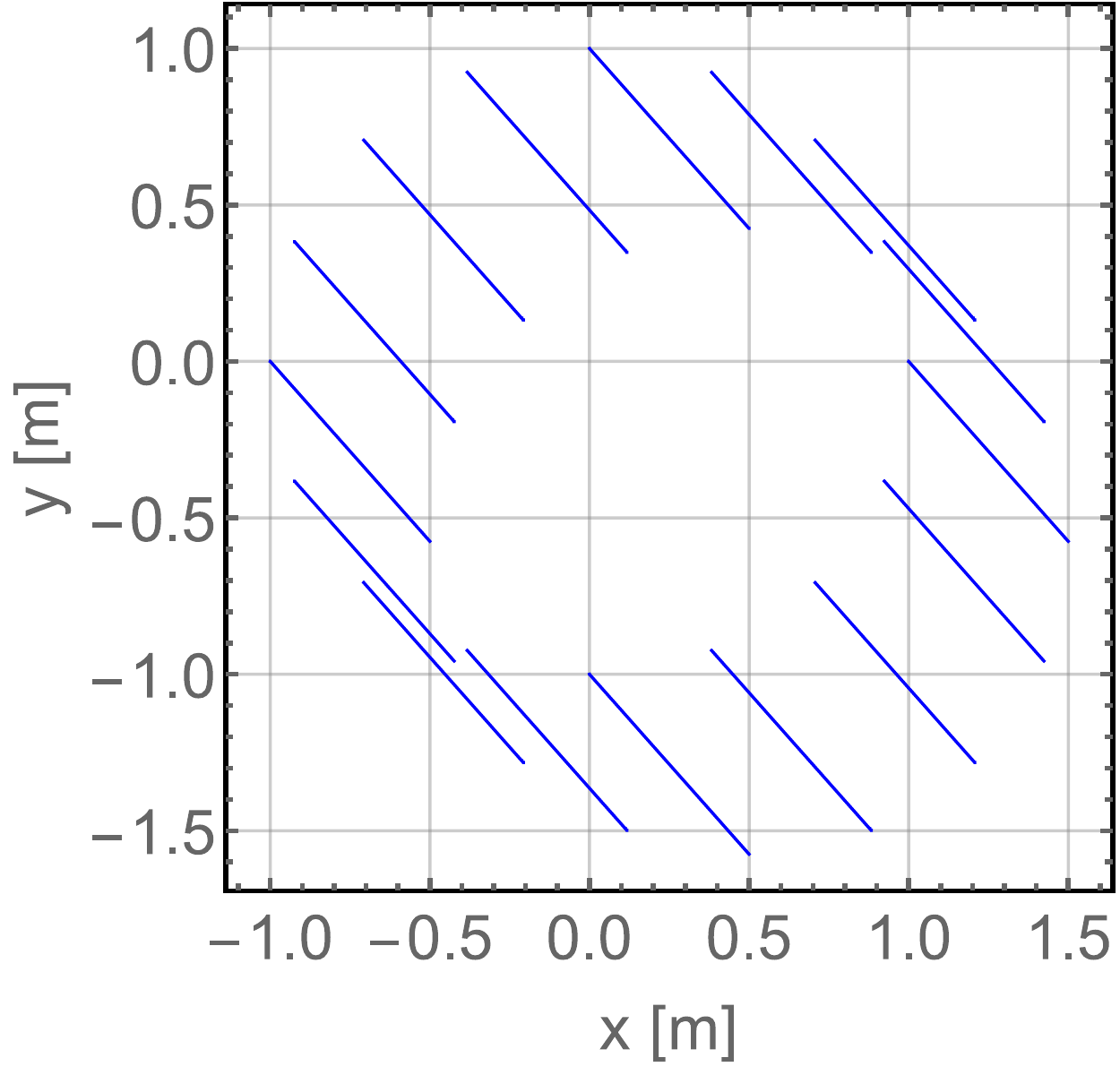}}
\caption{Ring consisting of individual, massive particles/test bodies including an interferometer whose mirrors can also be interpreted as test
bodies \protect\subref{fig:ring-of-particles}. Trajectories for particles sitting on a ring,
which is initially (for $\tau=0$) centered at the coordinate origin \protect\subref{fig:ring-plus-polarization},
\protect\subref{fig:ring-both-polarizations}. The particles are subject to a Lorentz-violating background field and they
are hit by a gravitational wave that is $(+)$ polarized \protect\subref{fig:ring-plus-polarization} and in a superposition
of both polarizations \protect\subref{fig:ring-both-polarizations}, respectively. All trajectories are obtained from
Eqs.~(\ref{eq:geodesic-equations-model}) based on the values in the first line of \tabref{tab:numerical-values}. In
\figref{fig:ring-both-polarizations} there are tiny oscillations that are difficult to spot at these scales.}
\label{fig:particle-ring-model-3}
\end{figure}

To better understand the implications of this motion on a realistic experiment consider a ring of such particles as shown in
\figref{fig:ring-of-particles}. This procedure is usually applied to study how a gravitational wave affects an extended object. Due to
Einstein's equivalence principle, in the standard theory the coordinates of a single particle do not change when it is hit by a
gravitational wave. What changes, though, is the distance between distinct particles separated from each other because the gravitational
wave induces a change of the metric. That effect will not be considered below but only the nonstandard motion of individual particles
that arises from the existence of the Lorentz-violating background field.

The motion of particles making up a ring, which follows from solving Eqs.~(\ref{eq:geodesic-equations-model}) numerically, is shown in
\figref{fig:particle-ring-model-3}. For a particular choice of polarization,
all particles move along congruent trajectories that differ from case to case, though. If the wave is (+) polarized each particle moves along the negative
$y$ direction only but along the $x$ direction it stays at the same coordinate. For the wave being $(\times)$ polarized the motion of a single particle
is more complicated. It moves both along the positive $x$-direction and the negative $y$-direction where the resulting motion is oscillatory. Last but
not least, for a superposition of both polarizations, the particles do not behave qualitatively differently from the case of a $(\times)$ polarized
wave.

\section{Impact on interferometer experiments}
\label{sec:observable-signals}

Now we are in a position to apply the model introduced above to the gravitational-wave signal detected by Advanced LIGO. Curvature effects at the surface of the
Earth, which originate from its Schwarzschild spacetime, have an order of magnitude of around $R_S/R_{\oplus}\approx 1.4\times 10^{-9}$ in comparison to
Minkowski spacetime. Here $R_S$ is the Schwarzschild radius of the Earth and $R_{\oplus}$ is the radius of the planet. Therefore, we will neglect these effects
on the Lorentz-violating background field and the propagation of the gravitational wave, i.e., the problem is still treated in a Minkowskian background
spacetime. Interferometric experiments such as LIGO are sensitive to motions in the plane of the apparatus, cf.~\figref{fig:particle-ring-model-3} for different
gravitational-wave polarizations. Suppose that an interferometric system is put into this coordinate system such that one arm points along the positive $x$-direction
and the other one along the positive $y$-direction, cf.~\figref{fig:ring-of-particles}. Since each particle moves in the same way, the whole ring, i.e., the entire
interferometer gets translated. In what follows, we will discuss how this motion can be detected, in principle.

\subsection{Interferometer response}
\label{sec:mirrors-in-phase}

The response of an interferometer whose arm lengths vary with time can be obtained from the round-trip distance that gives the pathlength traveled by
a wavefront back and forth a single arm of the interferometer. The wavefront shall be emitted at the position $y_1(0)$ of the first mirror at time $t=0$. It then
travels to the second mirror located at the time-dependent position $y_2(t)$ and shall hit this mirror at the time $t_a$, which follows from solving the equation
$y_2(t_a)-y_1(0)=ct_a$. The path length traveled is given by $l_1=y_2(t_a)-y_1(0)$. After being reflected by the second mirror, the wavefront propagates back to the
first mirror and hits it at the time $t_b$. The latter follows from $y_2(t_a)-y_1(t_b)=c(t_b-t_a)$. The path length covered along the way back is computed from
$l_2=y_2(t_a)-y_1(t_b)$. If the position of the mirrors does not change, the round-trip distance $\mathscr{L}_0$ is simply given by twice the arm length $L$. However,
since the mirrors move themselves the variation $\Delta\mathscr{L}$ of the round-trip distance is obtained from \cite{Bhawal:1998}
\begin{equation}
\Delta\mathscr{L}=l_1+l_2-\mathscr{L}_0=2y_2(t_a)-y_1(0)-y_1(t_b)-2L\,.
\end{equation}
This variation can be translated to a phase shift between the incoming wave and the wave reflected back. Therefore, once it is divided by the interferometer arm length
$L$ it can be interpreted as a gravitational-wave strain.

For $A_+$ and $\breve{a}_2/m_{\psi}$ small enough, it is possible to fit the numerical solution for the acceleration along the $y$ direction to an analytical
function $\ddot{y}(t)$. By integrating the latter twice with appropriate initial conditions, we obtain an approximation to the numerical solution $y(t)$.
Reinstating the speed of light $c$ leads to:
\begin{subequations}
\label{eq:approximations-solutions-motion}
\begin{align}
\ddot{y}(t)&\approx -\ddot{y}_0\sin(\omega t)\,,\quad \ddot{y}_0=\omega(1+A_+)\frac{c\breve{a}_2}{m_{\psi}}\,, \displaybreak[0]\\[2ex]
y(t)&\approx y_0\left[\sin(\omega t)-\omega t\right]\,,\quad y_0=\frac{\ddot{y}_0}{\omega^2}\,.
\end{align}
\end{subequations}
This was tested to be valid down to values $\breve{a}_2/m_{\psi}\ll 1$. The advantage of having an analytical function is that the influence of
numerical instabilities can be avoided that increase when solving the equations of motion for $\breve{a}_2/m_{\psi}\ll 1$. With the analytical
approximation $y(t)$ at hand, the characteristic times $t_a$ and $t_b$ are computed. The exact solutions are involved and not of much use, which is
why expansions in the dimensionless quantity $L\omega/c\ll 1$ are stated:
\begin{equation}
t_a=\frac{L}{c}-\frac{1}{6}\left(\frac{L\omega}{c}\right)^3\frac{y_0}{c}\,,\quad t_b=\frac{2L}{c}+\frac{7}{6}\left(\frac{L\omega}{c}\right)^3\frac{y_0}{c}\,.
\end{equation}
The round-trip variation normalized by the interferometer armlength gives a dimensionless number that mimics a gravitational-wave strain. At first
order in the previously used dimensionless quantity, it is given by
\begin{equation}
\frac{\Delta\mathscr{L}}{L}=\left(\frac{L\omega}{c}\right)^3\frac{y_0}{L}\,.
\end{equation}
A value of $y_0=\unit[7.6\times 10^{-9}]{m}$ leads to $\Delta\mathscr{L}/L\approx 1.1\times 10^{-18}$, which lies within the sensitivity of Advanced LIGO.
Inserting the duration $T_0$ of the gravitational-wave pulse into \eqref{eq:approximations-solutions-motion}, we obtain $|\breve{a}_2/m_{\psi}|=3.8\times 10^{-8}$
that would lead to such a displacement $y_0$.

\subsection{Time lag due to finite propagation speed}
\label{sec:mirrors-out-of-phase}

In principle, gravitational waves can come from all directions imaginable. Hence, a wave will influence one of the mirrors first and
the other after a short period. Since LIGO is a large-scale experiment, this time is not negligible, and at the maximum, it corresponds to
the time that a gravitational wave takes to propagate along the interferometer arm length $L$. Since gravity is taken to be conventional, the
gravitational wave travels with the speed of light, which results in the time lag $\Delta t\approx \unit[1.3\times 10^{-5}]{s}$.
So both mirrors move out of phase and there is a net displacement $\Delta y$ that is caused by the nonzero $\Delta t$. Let $t=0$ be the time
when the wavefront hits the second mirror. The displacement is then given by
\begin{equation}
\Delta y=|y^{(1)}(t)-y^{(2)}(t)|=|y(t+\Delta t)-y(t)|\,.
\end{equation}
For $\breve{a}_2/m_{\psi}=4.0\times 10^{-14}$ the maximum value for the displacement was found to be $\Delta y\approx \unit[1.1\times 10^{-18}]{m}$,
which lies within the sensitivity of Advanced LIGO.

\subsection{Macroscopic test bodies}

Until now, the Lorentz-violating background field was assumed to affect all particles in the same manner, regardless of their species.
However, this is not the case necessarily. Note that the CKM matrix in flavor physics is neither diagonal nor are their components equal. Instead, they
depend on the quark flavor. The spirit of the SME is similar, whereupon the controlling coefficients of the background field are taken to be dependent
on the species, in general. Hence, each particle species couples to Lorentz violation differently, which would make Lorentz-violating effects stronger
for one kind of particles compared
to another type. Now, the particle of mass $m_{\psi}$ will be considered as an extended test body with a realistic particle content. Each species
of particles shall couple to a different background field whose controlling coefficients are labeled appropriately.
The relevant coefficient for the test body $B$ is then \cite{Kostelecky:2010ze}
\begin{equation}
\label{eq:controlling-coefficient-decomposition}
\frac{\overline{a}^B_2}{m^B}=\sum_{w=e,p,n} \frac{N^w\overline{a}^w_2}{m^B}\,,\quad \frac{\breve{a}^B_2}{m^B}=\frac{\alpha}{2}A_+\sum_{w=e,p,n} \frac{N^w\overline{a}^w_2}{m^B}\,,
\end{equation}
where $m^B$ is the mass of the test body and $\overline{a}^w_2$, $\breve{a}^w_2$ are the controlling coefficients for electrons,
protons, and neutrons. Furthermore, $N^w$ are the numbers of the latter particles. Note that according to \eqref{eq:perturbation-coefficients},
$\overline{a}_2$ and $\breve{a}_2$ are linked. The fluctuation $\breve{a}_2$ is suppressed by the tiny amplitude
strain $A_+$ in comparison to $\overline{a}_2$.

\subsection{Mirrors of different molecular composition}
\label{sec:mirrors-composition}

Based on \eqref{eq:controlling-coefficient-decomposition}, let us imagine a hypothetical interferometric experiment involving two mirrors (1) and (2).
The first shall be composed of a material with $N^{(1)w}$ electrons, protons, and neutrons. The second is made up of a different material with
different numbers $N^{(2)w}$ of particles. The deformations of the background field for each mirror can then be written as follows:
\begin{equation}
\frac{\breve{a}^{(1)}_2}{M}=\frac{\alpha}{2}A_+\sum_{w=e,p,n} \frac{N^{(1)w}\overline{a}^w_2}{M}\,,\quad \frac{\breve{a}^{(2)}_2}{M}=\frac{\alpha}{2}A_+\sum_{w=e,p,n} \frac{N^{(2)w}\overline{a}^w_2}{M}\,,
\end{equation}
where the mass $M$ of both mirrors shall be equal. We again describe the solutions of the equations of motion approximately by
\eqref{eq:approximations-solutions-motion}. If the molecular composition of the mirrors is not the same they will be affected by Lorentz
violation differently. Hence, there would be a displacement $\Delta y$ between both mirrors that would change with time:
\begin{equation}
\Delta y=|y^{(1)}(t)-y^{(2)}(t)|=\frac{1}{\omega}(1+A_+)\Delta a|\omega t-\sin(\omega t)|\,,\quad \Delta a\equiv \left|\frac{\breve{a}_2^{(1)}}{M}-\frac{\breve{a}_2^{(2)}}{M}\right|\,.
\end{equation}
For the explicit values $A_+=10^{-21}$ and $\Delta a=10^{-17}$, which was found by evaluating $\Delta y$ for the duration of the impact
of the signal GW150914, the displacement
can reach $|\Delta y|\approx \unit[2.0\times 10^{-18}]{m}$. Therefore, Advanced LIGO would be sensitive to such a difference of the
controlling coefficients. This can be employed to derive a sensitivity on combinations of the controlling coefficients $\overline{a}_2^w$ for electrons,
protons, and neutrons.

\subsection{Experimental sensitivities}

\begin{table}[b]
\centering
\begin{tabular}{ccc}
\toprule
   & $N_p$ & $N_n$ \\
\midrule
H & 1 & 0 (99.9885\%), 1 (0.0115\%) \\
O & 8 & 8 (99.757\%), 9 (0.038\%), 10 (0.205\%) \\
Al & 13 & 14 (100\%) \\
Si & 14 & 14 (92.2297\%), 15 (4.6832\%), 16 (3.0872\%) \\
\bottomrule
\end{tabular}
\caption{Proton number $N_p$ and neutron number $N_n$ for hydrogen (H), oxygen (O), aluminum (Al), and silicon (Si). The numbers in parentheses
state the percentage of each stable isotope (see \cite{Elements:2016}).}
\label{tab:elements-date}
\end{table}
By and large, the test masses of the LIGO interferometer correspond to the two mirrors making up the Fabry-Perot cavity in $y$ direction.
Each mirror weighs \unit[40]{kg} and it largely consists of silica (SiO$_2$)~\cite{LIGO:2016}. Using the data of \tabref{tab:elements-date}
for the experimental setups described in Secs.~\ref{sec:mirrors-in-phase}, \ref{sec:mirrors-out-of-phase} we obtain the sensitivities in
the first two rows of \tabref{tab:experimental-sensitivity}. If the coupling constant $\alpha$ between gravity and the $a$ coefficients is of the
order of 1, the constant part $\overline{a}_2$ of the controlling coefficient can be enormous. This statement does not clash with flat-spacetime constraints
as the $a$ coefficients in Minkowski spacetime cannot be probed by a single fermion flavor, which allows such large values, in principle
\cite{Kostelecky:2008in}. Unfortunately, the first sensitivity in \tabref{tab:experimental-sensitivity} exceeds the bounds obtained in~\cite{Kostelecky:2010ze}
by many orders of magnitude. The bounds in the latter paper are based on experiments with test masses in the gravitational field of Earth such as free-fall
tests and satellite-based tests of the weak equivalence principle (see, e.g., Tab. VI, VIII, and XI in \cite{Kostelecky:2010ze}).
The large gravitational field of the Earth (in comparison to a gravitational-wave perturbation) flows into these constraints.
Furthermore, a change of the round-trip variation in the LIGO interferometer is suppressed by the third power of the small quantity $L\omega/c$.
Both facts render the sensitivity of LIGO poor for this particular kind of Lorentz violation.

Taking into account the curvature of Earth's surface, the mirrors at the end of the interferometer arm would move out of phase, which can be
exploited. It improves the sensitivity by many orders of magnitude, and it leads to the number in the second row of \tabref{tab:experimental-sensitivity}.
The result still lies many orders of magnitude above the current
sensitivity of experiments in the gravitational field of the Earth, though. Last but not least, the sensitivity can be slightly improved again in
a hypothetical gravitational interferometer whose mirrors consist of different materials,
cf.~\secref{sec:mirrors-composition}. We assume that one of the mirrors is composed of $\mathrm{SiO_2}$ such as in the LIGO experiment.
In principle, such a mirror must be constructed from a crystalline material with a high mechanical quality factor.
Since in the design phase of Advanced LIGO it was planned to install sapphire mirrors we choose corundum $\mathrm{Al_2O_3}$
as a hypothetical material for the second mirror.

This pair of mirrors would move out of phase when Lorentz violation does not couple to different particle species in a
uniform way. Such a flavor-dependent scenario of Lorentz violation is realistic, as described above. The data stated in
\tabref{tab:elements-date} enables us to derive the sensitivity given in the third row of \tabref{tab:experimental-sensitivity}.
In contrast to the previous two sensitivities, the current one has a minus sign in front of the neutron number, which shows
that it relies on the difference between the numbers of protons/electrons and neutrons. 
This is the best sensitivity for Lorentz violation in the fermion sector obtained in
the context of interferometric gravitational-wave experiments. In general, the sensitivity improves whenever we do not have to
rely on the measurement of an indirect quantity that is heavily suppressed such as the round-trip variation. Instead, in the experimental
setups of Secs.~\ref{sec:mirrors-out-of-phase}, \ref{sec:mirrors-composition}, Lorentz violation can be looked for directly by
measuring the displacement of the mirrors.
\begin{table}[t]
\centering
\begin{tabular}{ccl}
\toprule
Experimental setup & Combination of coefficients & \multicolumn{1}{c}{Sensitivity} \\
\midrule
Direct interferometric response, Sec.~\ref{sec:mirrors-in-phase} & $|\alpha(\overline{a}^e_2+\overline{a}^p_2)+\alpha\overline{a}^n_2|$ & $\unit[1.4\times 10^{14}]{GeV}$ \\
Maximum propagation time lag, Sec.~\ref{sec:mirrors-out-of-phase} & $|\alpha(\overline{a}^e_2+\overline{a}^p_2)+\alpha\overline{a}^n_2|$ & $\unit[1.5\times 10^8]{GeV}$ \\
Different composition of mirrors, Sec.~\ref{sec:mirrors-composition} & $|\alpha(\overline{a}^e_2+\overline{a}^p_2)-\alpha\overline{a}^n_2|$ & $\unit[2.1\times 10^6]{GeV}$ \\
Satellite Doppler tracking, Sec.~\ref{sec:beyond-ligo} & $|\alpha(\overline{a}^e_2+\overline{a}^p_2)+1.1\alpha\overline{a}^n_2|$ & $\unit[2.9\times 10^{15}]{GeV}$ \\
\bottomrule
\end{tabular}
\caption{Experimental sensitivities for combinations of flavor-dependent $\overline{a}_2$ based on different interferometric setups.}
\label{tab:experimental-sensitivity}
\end{table}

\subsection{Experiments beyond LIGO}
\label{sec:beyond-ligo}

One might think that better sensitivities may be obtained by other experiments such as pulsar timing measurements \cite{Hobbs:2009yy} and
satellite Doppler tracking \cite{Armstrong:2006zz}. The first kind of measurements is carried out by the International Pulsar Timing Array.
This project includes a collection of telescopes that observe arrival times for radio pulses emitted by pulsars. The strain of
a gravitational wave may induce a frequency change on such pulses, which would affect their arrival time. The second type of experiment uses the
observation that a gravitational wave can interact with propagating radio waves, which results in an apparent Doppler shift of the wave. The
latter shift is related to the difference of the strain amplitudes at the emitter and detector of the radio pulse, which offers the possibility
of detecting gravitational waves. Hence, these categories of experiments rely on the detection of radio pulses or the evaluation of pulse shape curves,
i.e., they can be expected to enable searches for Lorentz violation in the photon and gravity sector primarily. However, in
principle they could be employed for our purpose as well, i.e., to study the impact of Lorentz violation in the fermion sector on such
measurements.

One could imagine a gravitational wave to hit both the Earth and the satellite at the same time. This happens when the wave vector
is perpendicular to the vector pointing from the center of the Earth to the satellite. In such a situation the system can be interpreted as an
interferometer with a very large arm length resulting in a much better sensitivity than that provided by Advanced LIGO. However, such a situation
is fine-tuned and unrealistic, which is why it will not be considered. Instead, we assume that the wave hits the satellite first at which point the
satellite would start moving with an average velocity of magnitude $|\ddot{y}_0/\omega|$, cf.~\eqref{eq:approximations-solutions-motion}. Such a
motion would induce a frequency variation $\delta\nu/\nu$ due to the relativistic Doppler effect where $\nu$ is the frequency of the signal sent
from the Earth to the satellite and back. As the signal is sent back and forth, it is affected twice. A maximum velocity of
$|\dot{y}|=\unit[1.5\times 10^{-6}]{m/s}$, which is connected to $|\breve{a}_2/m_{\psi}|=7.5\times 10^{-7}$, would
lead to frequency variations of $|\delta \nu/\nu|=10^{-14}$. This value is taken as a conservative sensitivity of those experiments as all
noise is estimated to lie at the level of $10^{-15}$ and below \cite{Armstrong:2006zz}. To get
a crude estimate of the sensitivity to detecting Lorentz violation in the fermion sector, the satellite is considered as a test body made of
aluminum. The result is stated in the fourth line of \tabref{tab:experimental-sensitivity}. It is around one order of magnitude worse than
that obtained from the interferometer response of Advanced LIGO.

\section{Conclusions and outlook}
\label{sec:conclusion}

In this paper, a simple model for a classical particle subject to Lorentz violation was the focus. The model was based on the
point-particle Lagrangian that is linked to the $a$ coefficients of the minimal Standard-Model Extension (SME). In light of the
recent direct detection of gravitational waves by Advanced LIGO, the particle was coupled to a gravitational-wave metric perturbation.
By doing so, the leading-order geodesic equations were obtained and solved numerically. In contrast to the Lorentz-invariant
theory, a single particle exhibits a (periodic) accelerated motion caused by combined effects of the gravitational wave and the
Lorentz-violating background field. The numerical solutions were used to compute sensitivities for different experimental
setups: (1) the direct interferometric response of Advanced LIGO caused by a variation of the round-trip distance,
(2) considering the maximum time lag between the interaction of the gravitational wave with the mirrors, (3) a hypothetical
interferometric experiment whose mirrors have distinct molecular compositions, and (4) satellite Doppler tracking, which is a
method for detecting gravitational waves beyond LIGO. The first two setups give a range of sensitivities
$\unit[1.5\times 10^8]{GeV}< \Xi<\unit[1.8\times 10^{16}]{GeV}$ on a particular combination $\Xi$ of controlling coefficients for
protons, neutrons, and electrons. The minimum follows from the situation when the wavefronts are perpendicular to the vector
pointing from one mirror to the other. The maximum holds when the wavefronts are parallel to the vector connecting both mirrors
such that the sensitivity based on the interferometer response plays the dominant role. By mischance, these two sensitivities lie many
orders of magnitude above
current constraints from tests of the weak-equivalence principle on Earth and satellites. The poor sensitivity can be traced back to the extremely weak field
of a gravitational wave and, partially, due to further suppressions of the observable signal by the small quantity $L\omega/c$.
The third sensitivity that could be achieved with the hypothetical interferometric setup is slightly better than the second because species-dependent Lorentz
violation would bring the two mirrors out of phase directly. The fourth sensitivity that could be obtained from satellite
Doppler tracking experiments is slightly worse than the first. Hence, the results of the paper should not be considered in
the light of having new impressive constraints on Lorentz violation. Instead, the sensitivities here in combination with the established
constraints from weak-equivalence principle tests demonstrate that the impact of Lorentz violation in the fermion sector is negligible
in gravitational-wave detection by interferometric or other means.

This paper can be the base for further applications of point-particle Lagrangians for the SME fermion sector in the context of
gravitational-wave physics. As already mentioned, the phenomenological approach discussed here is not suitable to obtain constraints
on fermionic controlling coefficients. However, there is a plethora of theoretical questions to be answered. First, for explicit
Lorentz violation, the implications of a nonconserved energy-momentum tensor at the source for gravitational waves will be a worthwhile
issue to investigate. We will then explore whether and how the clash with the geometrical Bianchi identities can be circumvented within
the realm of Finsler geometry (cf.~\cite{Schreck:2015dsa} for preliminary studies in the photon sector). Second, models based on
alternative Lagrangians for fermions can be explored to gain more theoretical insight, e.g., the bipartite case of the $b$ coefficients
\cite{Kostelecky:2010hs,Kostelecky:2011qz,Kostelecky:2012ac}.
It will also be interesting to find out if for such alternative sets of coefficients the geodesic equations for spontaneous Lorentz
violation and explicit Lorentz violation in the realm of Finsler geometry correspond to each other at first order in Lorentz violation
(cf.~\ref{sec:remarks-finsler-geometry}). By doing so, we could find out whether this is a general property or specific for the $a$ coefficients only.

\section{Acknowledgments}

It is a pleasure to thank V.A.~Kosteleck\'{y} for helpful discussions on the manuscript and the concepts of spontaneous and
explicit Lorentz violation. Furthermore, the author acknowledges most useful suggestions by the three
anonymous referees that made the focus and the structure of the paper more appropriate. This work was partially funded by the Brazilian
foundation FAPEMA.

\begin{appendix}
\setcounter{equation}{0}
\newpage

\section{Remarks on explicit Lorentz violation and Finsler geometry}
\label{sec:remarks-finsler-geometry}

As was mentioned in \secref{sec:models}, explicit Lorentz violation in a curved gravitational background clashes with some of the
basic properties of Riemannian geometry such as the Bianchi identities. Therefore, a reasonable alternative to spontaneous Lorentz
violation could be to resort to a more general geometrical framework such as Finsler geometry. To do so we have to propose an
{\em Ansatz} for an explicitly Lorentz-violating vector field. It is a reasonable assumption for this background field to adopt at
least some of the symmetries or properties of the curved spacetime where it is defined on. For example, in a curved, isotropic spacetime
such as the Schwarzschild metric it is rational to assume that a Lorentz-violating background field is spherically symmetric as well,
cf.~\cite{Schreck:2015dsa}. Therefore, when the gravitational wave propagates through empty space the background field shall adapt
to the metric perturbation caused by the wave. One possibility is an $a_{\mu}$ with $a_{\mu}k^{\mu}=0$, which is periodic in $k\cdot X$:
\begin{subequations}
\begin{align}
\label{eq:general-choice-background-vector}
a_{\mu}(k\cdot X)&=\overline{a}_{\mu}+\breve{a}_{\mu}\cos(k\cdot X)\,, \\[2ex]
\breve{a}_{\mu}&=\begin{pmatrix}
0 \\
f_1(A_+,A_{\times}) \\
f_2(A_+,A_{\times}) \\
0 \\
\end{pmatrix}_{\mu}\,,
\end{align}
\end{subequations}
with unknown functions $f_{1,2}=f_{1,2}(A_+,A_{\times})$ of the strain amplitudes. The latter $a_{\mu}$ has a form very similar
to \eqref{eq:choice-background-field-model}. However, its interpretation is completely different. Here, both $\breve{a}_1$ and
$\breve{a}_2$ are put in by hand and, therefore, they are nondynamical. Besides, they are {\em a prori} not linked to the strain
amplitudes $A_+$, $A_{\times}$ of the gravitational wave. Hence, the functions $f_{1,2}$ were introduced (also by hand) to account for the
fact that the perturbation vanishes when there is no gravitational wave. Therefore, $f_{1,2}(0,0)=0$ should hold. A first-order
linearization will be $f_{1,2}(A_+,A_{\times})\approx \xi_{1,2}A_++\zeta_{1,2}A_{\times}$ with unknown constants $\xi_{1,2}$,
$\zeta_{1,2}$.

In principle, \eqref{eq:general-choice-background-vector} could be interpreted as the leading terms of a Fourier decomposition of a
periodic $a_{\mu}(k\cdot X)$ with the higher-order harmonics neglected. Hence, the controlling coefficients shall exhibit an oscillatory
behavior such as the wave does.
The leading sine contribution in its Fourier expansion is supposed to vanish. A nonzero phase shift in the argument of the cosine function could be introduced to
model a delayed response of the Lorentz-violating background field to the metric perturbation. However, the response shall take place
instantaneously setting the phase shift to zero. This whole argument shows that for explicit Lorentz violation a range of assumptions
has to be made on how the Lorentz-violating vector field could look like. For spontaneous Lorentz violation, the form of $a_{\mu}$
follows directly from field equations.

Next, we consider $\widetilde{L}^{(a)}$ of \eqref{eq:lagrangian-minimally-coupled} with an explicitly Lorentz-violating vector
field $a_{\mu}$ inserted. A Wick rotation-type procedure then makes $\widetilde{L}^{(a)}$ a
Riemann-Finsler structure \cite{Kostelecky:2011qz,Kostelecky:2012ac}. For a textbook treatment of Finsler geometry, we refer to
\cite{Antonelli:1993,Bao:2000}. Since $\widetilde{L}^{(a)}$ involves a pseudo-Riemannian metric $g_{\mu\nu}$ it must be treated as
a pseudo-Riemann-Finsler structure, i.e., as the integrand of a generalized path length functional on a curved spacetime with an
intrinsic preferred direction. The knowledge of such Finsler spacetimes is not as elaborate as that of Finsler spaces. Certain properties
of Finsler spacetimes, e.g., causality are quite subtle. Nevertheless, Finsler spacetimes were studied in the series of papers \cite{Pfeifer:2011tk,Pfeifer:2011ve,Pfeifer:2012mb}, amongst others. Besides, the recent articles \cite{Silva:2015ptj,Silva:2016qkj}
are devoted to field theories in a Finsler spacetime characterized by \eqref{eq:lagrangian-minimally-coupled}. In the current section,
we will not delve into formal issues with regards to Finsler spacetimes. A reasonable expectation is that for the components of
$a_{\mu}$ small enough and for the particle velocity far away from the speed of light the usual relations used for a Finsler
structure can also be applied to this particular case of a pseudo-Finsler structure.

In a general Finsler space, the Riemannian metric $g_{\mu\nu}$ is used to compute lengths of vectors and enclosed angles in its
tangent space. Any Finsler space has an additional metric associated with it, which is called the Finsler metric $\widetilde{g}_{\mu\nu}$:
\begin{equation}
\label{eq:finsler-metric}
\widetilde{g}_{\mu\nu}\equiv\frac{1}{2}\frac{\partial (\widetilde{L}^{(a)})^2}{\partial u^{\mu}\partial u^{\nu}}\,.
\end{equation}
Note that the latter is denoted with a tilde to distinguish it from the (pseudo)-Riemannian metric $g_{\mu\nu}$.
The Finsler metric allows for computing the Finsler equivalent of the affine connection, i.e., the Christoffel symbols. They are
defined such as in Riemannian geometry but they now involve the Finsler metric:
\begin{equation}
\label{eq:christoffel-symbols}
\widetilde{\gamma}^{\mu}_{\phantom{\mu}\nu\varrho}\equiv\frac{1}{2}\widetilde{g}^{\mu\sigma}\left(\frac{\partial \widetilde{g}_{\sigma\nu}}{\partial x^{\varrho}}+\frac{\partial \widetilde{g}_{\sigma\varrho}}{\partial x^{\nu}}-\frac{\partial \widetilde{g}_{\nu\varrho}}{\partial x^{\sigma}}\right)\,.
\end{equation}
We do not consider any torsion effects, which makes the Christoffel symbols symmetric in the lower pair of indices:
$\widetilde{\gamma}^{\mu}_{\phantom{\mu}\nu\varrho}=\widetilde{\gamma}^{\mu}_{\phantom{\mu}\varrho\nu}$. In Finsler geometry,
the usual geodesic equations are generalized with the Finslerian affine connection. In general, for a geodesic
$x^{\mu}=x^{\mu}(\lambda)$ that is parameterized using an arbitrary parameter $\lambda$ they read
\begin{equation}
\label{eq:geodesic-equations-general}
\widetilde{L}^{(a)}\frac{\mathrm{d}}{\mathrm{d}\lambda}\left(\frac{1}{\widetilde{L}^{(a)}}\frac{\mathrm{d}x^{\mu}}{\mathrm{d}\lambda}\right)+\widetilde{\gamma}^{\mu}_{\phantom{\mu}\nu\varrho}\frac{\mathrm{d}x^{\nu}}{\mathrm{d}\lambda}\frac{\mathrm{d}x^{\varrho}}{\mathrm{d}\lambda}=0\,.
\end{equation}
Note that in the majority of the literature on (Finsler) geometry, geodesics are assumed to be parameterized such that the speed
of a particle traveling along the geodesic remains constant. Hence, the Finsler structure, which is the integrand of the path length
functional and, therefore, can be understood as a generalized speed, does not change. The interpretation is different for a
Lagrangian, which is a pseudo-Finsler structure. Nevertheless, there are parameterizations leaving $L^{(a)}$ unchanged along the
geodesic, which renders the first term in \eqref{eq:geodesic-equations-general} the usual four-acceleration. In the Lorentz-invariant
case such a parameterization is possible using proper time. However, when there are Lorentz-violating contributions a suitable choice
is more involved. So even if we still parameterize a geodesic with proper time, which is what we want to do, it is not granted that
this is a constant-speed parameterization. Hence, the geodesic equations will be taken as they stand in
\eqref{eq:geodesic-equations-general}.

Based on the Lagrangian of \eqref{eq:lagrangian-minimally-coupled}, both the Finsler metric and the Finsler affine connection
coefficients can be computed for an $a_{\mu}$ with $a_1\neq 0$ and $a_2\neq 0$. It is reasonable to carry out all computations
with computer algebra. The exact results are lengthy and do not provide any further insight into the problem, which is why they
will be skipped. Finally, they are used to obtain the geodesic equations based on \eqref{eq:geodesic-equations-general}. The full
equations are also very lengthy. However, an expansion at first order in the velocities, the amplitude strains, and products of
the amplitude strains with $a_1$, $a_2$ is reasonably compact enough to be stated:
\begin{subequations}
\label{eq:geodesic-equations-model-finsler}
\begin{align}
\label{eq:geodesic-equation-t-component-finsler}
0&=\ddot{t}-\ddot{x}\frac{a_1}{m_{\psi}}-\ddot{y}\frac{a_2}{m_{\psi}}\,-\,\ddot{z}\frac{a_3}{m_{\psi}} \notag \\
&\phantom{{}={}}+\left(\dot{y}\frac{a_2A_+-a_1A_{\times}+a_2'}{m_{\psi}}-\dot{x}\frac{a_1A_{+}+a_2A_{\times}-a_1'}{m_{\psi}}\right)\omega\sin(k\cdot X)\,, \displaybreak[0]\\[2ex]
\label{eq:geodesic-equation-x-component-finsler}
0&=\ddot{x}+\left[A_+\dot{x}+A_{\times}\dot{y}+(1-\dot{z})\frac{a_1'}{m_{\psi}}\right]\omega\sin(k\cdot X)\,, \displaybreak[0]\\[2ex]
\label{eq:geodesic-equation-y-component-finsler}
0&=\ddot{y}+\left[A_{\times}\dot{x}-A_+\dot{y}+(1-\dot{z})\frac{a_2'}{m_{\psi}}\right]\omega\sin(k\cdot X)\,, \displaybreak[0]\\[2ex]
\label{eq:geodesic-equation-z-component-finsler}
0&=\ddot{z}+\left(\dot{x}\frac{a_1'}{m_{\psi}}+\dot{y}\frac{a_2'}{m_{\psi}}\right)\omega\sin(k\cdot X)\,.
\end{align}
\end{subequations}
Comparing these relationships to Eqs.~(\ref{eq:geodesic-equations-model}) reveals that they almost correspond to each other. The only
differences are the contributions that involve $a_i$ directly and not their derivatives. However,
Eqs.~(\ref{eq:geodesic-equation-x-component-finsler}), (\ref{eq:geodesic-equation-y-component-finsler}), and  (\ref{eq:geodesic-equation-z-component-finsler})
can be solved for the accelerations and be inserted into \eqref{eq:geodesic-equation-t-component-finsler}, which eliminates all terms
dependent on $a_i$. Terms that involve products $a_ia_j'$ are neglected since those are of higher order.

Note that Eqs.~(\ref{eq:geodesic-equations-model-finsler}), as they stand, are valid for a general explicitly Lorentz-violating
vector field $a_{\mu}$ with nonzero components $a_1$, $a_2$, i.e., the particular \textit{Ansatz} of \eqref{eq:general-choice-background-vector}
has not been employed so far. It is an interesting observation that the geodesic equations for spontaneous and explicit Lorentz violation
treated in the realm of Finsler geometry are equal at leading order in Lorentz violation and strain. Choosing
$\xi_{1,2}=(-1)^{1,2}(\alpha/2)\overline{a}_{1,2}$ and $\zeta_{1,2}=-(\alpha/2)\overline{a}_{2,1}$
the \textit{Ansatz} of \eqref{eq:general-choice-background-vector} corresponds to \eqref{eq:choice-background-field-model}.
We emphasize again that both vector fields have entirely different interpretations. The choice of \eqref{eq:choice-background-field-model}
is obtained from a field equation for $a_{\mu}$, whereas all properties of \eqref{eq:general-choice-background-vector} must be put in by hand.

In the context of spontaneous Lorentz violation, the energy-momentum tensor is covariantly conserved on-shell, i.e., when the geodesic
equations (\ref{eq:geodesic-equations-model}) for the particle and the field equations for the Lorentz-violating background field are
fulfilled. The geodesic equations (\ref{eq:geodesic-equations-model-finsler}) for explicit Lorentz violation correspond to
Eqs.~(\ref{eq:geodesic-equations-model}) at first order in Lorentz violation. So the lately mentioned choice for the explicitly Lorentz-violating
vector field is an approximation to the $a_{\mu}$ obtained from spontaneous symmetry breaking. The energy-momentum tensor must be covariantly
conserved for solutions of the geodesic equations based on this particular explicitly Lorentz-violating $a_{\mu}$. The latter must then
also be consistent with the Bianchi identities at first order in Lorentz violation. We will elaborate on theoretical questions of this
and of a similar kind in future work.

\end{appendix}

\newpage


\end{document}